\documentclass[%
superscriptaddress,
 amsmath,amssymb,
 aps,
onecolumn,
tightenlines,
11pt]{revtex4-2}

\usepackage{graphicx}
\usepackage{dcolumn}
\usepackage{bm}
\usepackage{xcolor}
\usepackage{appendix}
\usepackage{hyperref}
\usepackage{cleveref}

\begin{document}

\preprint{APS/123-QED}

\title{Zeros of the partition function for 12 flavor QCD}

\author{Anas Saleh}
\email{(affiliated at the time the work was completed)}
\affiliation{Department of Physics and Astronomy, University of Iowa\\30 N Dubuque St, Iowa City, IA 52242}
\author{Michael Hite}
\email{michaelhite@arizona.edu}
\affiliation{Department of Physics, University of Arizona\\
 1118 E 4th St, Tucson, AZ 85719}
\author{Diego Floor}
\email{(affiliated at the time the work was completed)}
\affiliation{Department of Physics and Astronomy, University of Iowa\\30 N Dubuque St, Iowa City, IA 52242}
\author{Yannick Meurice}
\email{yannick-meurice@uiowa.edu}
\affiliation{Department of Physics and Astronomy, University of Iowa\\30 N Dubuque St, Iowa City, IA 52242}


\date{\today}

\begin{abstract}
We consider a four dimensional $SU(3)$ lattice gauge theory with 12 staggered fermions having identical masses and an unimproved action. Using sets of plaquette distributions for various inverse bare couplings $\beta$, we reconstruct the density of states with the Ferrenberg -Swendsen method and calculate the zeros of the partition in the complex $\beta$ plane with bare quark masses $m_q$ = 0.02, 0.06, 0.08 and 0.1 for hypercubes of linear size $L$= 4, 6, 8, 10, and 12. Our hypothesis is that there is a line of first order transitions in the $(m_q,\beta)$ plane ending at a second order phase transition. We expect this transition to be in the 4D Ising, mean field, universality class. We fit the $L$ dependence of the zeros with the lowest imaginary part using two ($y = bL^{-d}$) and  three ($y = a + bL^{-d}$) parameter fits. For $m_q$ = 0.02 the results provide strong support for a first order phase transition ($d=3.98(6)$, and $a$ statistically compatible with 0). The results also indicate, with less statistical significance for $m_q=0.06$, that the three other masses are above the critical value $m_q^c$. In addition, we suggest that the infinite volume gap for the lowest zero $a$, can be represented as $a\simeq A(m_q-m_q^c)^{B}$ with $m_q^c\sim 0.05$ and $B\sim 1$. Given that there are only three data points with significant error bars, it is difficult to rule out the mean field value $B=3/2$. Combining this result with spectroscopic results by Jin and Mawhinney, indicates that the gap with real axis (Lee-Yang edge) scales roughly like $m_\sigma ^2$, where $m_\sigma $ is the mass of the $0^{++}$ scalar which is also the lowest excitation.
\end{abstract}

\maketitle


\section{Introduction}
The experimental discovery of the Brout-Englert-Higgs boson \cite{ATLAS:2012yve,CMS:2012qbp} represents the culmination of theoretical ideas used to generate the masses of the W, Z and quarks in the standard model. Despite these successes, the analogy with superconductivity and arguments including naturalness reviewed in Ref. \cite{Craig:2022eqo} have led researchers to keep considering the hypothesis that the scalar sector of the standard model could be part of an effective theory coming from a higher energy scale, asymptotically free, gauge theory. In this context, a SU(3) gauge theory with $N_f=12$ fundamental flavors is quite interesting and its low energy effective theory differs significantly \cite{Jin:2009mc,Jin:2011eon,Fodor:2011tu,Appelquist:2011dp,Deuzeman:2011pa,Aoki:2012eq,Kuti:2014epa,DeGrand:2015zxa} from the one associated with Quantum Chromodynamics (QCD) with two or three light flavors. $N_f=12$ SU(3) is a strongly interacting theory and its massless limit is non trivial. Various points of views regarding the massless limit (conformal versus confining) are discussed in Refs. \cite{Hasenfratz:2024fad,LatKMI:2025kti,Fodor:2017gtj}. Recently, for $N_f<12$, Ref. \cite{klinger2026phasestructuremasslessmanyflavour} showed that the chiral transition in continuum QCD is of second order for all $N_f$ up to the onset of the conformal window.

Numerical studies of the massive $N_f=12$ SU(3) theory show that the mass of the $0^{++}$, called $\sigma$ or $f_0$ in the conventional QCD context,  can be smaller than the masses of the pions \cite{Jin:2009mc,Jin:2011eon,Cheng:2011ic,Fodor:2011tu,Jin:2013hpa,Aoki:2012eq,LatKMI:2013bhp,LatKMI:2025kti}. In Refs. \cite{Jin:2011eon,Jin:2013hpa}, it was observed that by increasing the bare quark mass $m_q$, it was possible to reach a critical point at the end of a line of first order phase transition in the $(\beta ,m_q)$ plane, where $\beta=6/g^2$ is associated with the gauge coupling $g^2$. Furthermore, it was found in Refs. \cite{Jin:2011eon,Jin:2013hpa} that $m_\sigma \propto (m_q-m_q^c)^{1/2}$, which seems compatible with a mean field second order phase transition as expected for the four-dimensional Ising model. This critical point was interpreted as a lattice feature that could be important to understand the distortion of the continuum limit for vanishing mass. It should be noted that effective linear sigma models with determinant terms designed to describe the effects of topological configurations \cite{PhysRevD.21.3388,kawarabayashi_ohta_1980,Meuricea0} contain terms that contribute to the scalar and pseudo scalar singlets with opposite signs. In tree level mass formulas, the lowering of the scalar mass gets more pronounced as the the number of flavor increases \cite{Meurice:2017zng,Floor:2018ytw}. It is thus conceivable, that these effective interactions could be able to generate scalar instabilities in the continuous effective theory independently of the lattice regularization. 

\begin{figure}[h]
    \centering
    \includegraphics[width=0.8\linewidth]{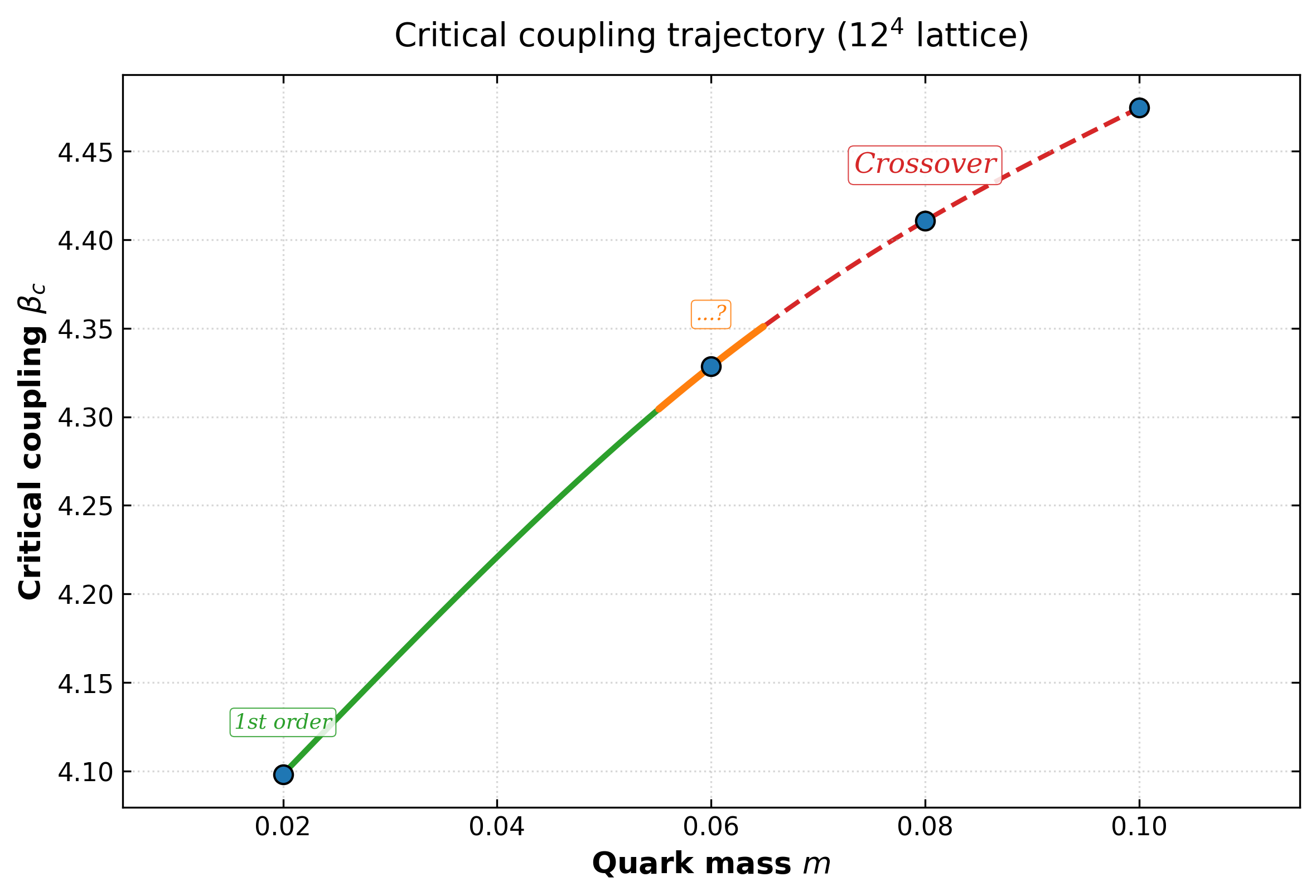}
    \caption{Expected phase diagram for 12 flavor $SU(3)$ in the $(m_q,\beta)$ plane. The critical values of $m_q$ and $\beta$ are estimated in Sec. \ref{sec:results} using the Fisher zeroes.}
    \label{fig:phase-behavior}
\end{figure}

In this article, we discuss the critical point using the finite-size scaling of the zeros of the partition function for the microscopic theory with four different bare masses $m_q$ = 0.02, 0.04, 0.06 and 0.1. Our general expectation is that there is a clear first order phase transition for the lowest mass (0.02) and a clear crossover for the largest mass (0.1). Our hypothesis is that 
a line of first order transition in the $(m_q,\beta)$ plane ends at a second order phase transition as illustrated in Fig. \ref{fig:phase-behavior}.

Spectroscopic results \cite{Jin:2009mc,Jin:2011eon,Jin:2013hpa} suggest that the $0^{++}$, a real scalar field, becomes massless at the second-order transition with an approximate scaling $m_\sigma \propto (m_q-m_q^c)^{1/2}$. This could correspond to the Ising universality class which is mean field in 4 dimensions and has $\nu=1/2$. For the Ising model with a given temperature $T$ and external magnetic field $h$, these two variables control the distance from the fixed point along the two relevant directions. They are ``orthogonal" because $h=0$ corresponds to the absence of explicit breaking of the $Z_2$ symmetry. For the 12 flavor model considered here, $m_q$ plays a role analogous to $T$ in the Ising model. For $m_q>m_c$, there is no discontinuity in the average gauge action as we vary $\beta$, but a discontinuity appears at $\beta_c(m_q)$ for $m_q<m_q^c$. However, the two directions are not ``orthogonal" and $\beta_c(m_q)$ is a nontrivial function (Fig. \ref{fig:phase-behavior}). 

In the following, we consider the 12 flavor SU(3) gauge theory with an unimproved action in the staggered formalism. Partial results have already been presented in \cite{gelzer2013fisherzerosrgflows, gelzer2014exploringphasestructure12flavor, e2018critical}. We perform a more detailed analysis of the Fisher zeros and the error bars, and compare with previous results. The main differences with Ref. \cite{e2018critical} are in the method used to calculate the relaxation time, the criteria for dropping the thermalization period, and the error estimation and bootstrapping. After calculating the Fisher zeros with error bars, we fit their imaginary part as a function of the linear size of the system for each of the masses. Having data for five different linear sizes, we perform fits with two and three parameters and compare their reduced $\chi^2$ and $p$-value. 

We expect to be in the vicinity of a second order transition of the Ising type controlled by $m_q$. For $m_q<m_q^c$, we have a first order transition as we vary $\beta$ and we expect that $Im \beta (L) \propto L^{-D}$ with $D=4$ \cite{janke2000}. For $m_q=m_q^c$, we have a second order transition as we vary $\beta$ and we expect that $Im \beta (L) \propto L^{-1/\nu}$ with $\nu=1/2$ \cite{itzykson83}. For  $m>m_q^c$ the two phases are continuously connected and the work of Lee and Yang \cite{leeyang} suggests that the zeros don't pinch the real axis. 

The article is organized as follows. In Sec. \ref{Model}, we introduce the lattice action. In Sec. \ref{scaling}, we review the general scaling expectations. We then discuss our methods for determining the zeros of the partition function and fitting procedures in Sec. \ref{allresults}. Finally, we interpret the implications of our results.

\section{Model}\label{Model}
We begin with the discretized QCD action whose gauge group is SU(3). The lattice has dimensions $3+1$ and volume $V=L_x^3\times L_t=L^4$. The fields $\psi_x, \bar{\psi}_x$ are defined on the lattice sites. The fermionic part of the lattice action is
\begin{equation}
    S_F = \frac{1}{2}\sum_x \bar{\psi}_x \left(\gamma_\mu\Delta^\mu + m_q\right)\psi_x + h.c.,
\end{equation}
where $m_q$ is the quark mass, $\gamma_\mu\Delta^\mu$ is the lattice Dirac operator, and the field variables include an implicit sum over flavors. The gauge part of the action is given by the Wilson action
\begin{equation}
    S_G = \beta\sum_p \left[1 - \frac{1}{3}\text{Re}\left(\text{Tr}U_p\right)\right]
\end{equation}
where $\beta=6/g^2$, $U_p$ is the product of the gauge fields on a single plaquette $p$. To resolve the doubling problem we use Kogut-Susskind's staggered formulation \cite{kogut-susskind}. Our observable of interest is the average plaquette value $E=\langle 1-\frac{1}{3}\text{Tr}U\rangle$ over the lattice for a given $(L,m_q,\beta)$. The average plaquette value is then used to build the partition function, written in terms of the density of states $\Omega(E)$ as
\begin{equation}
    Z(\beta) = \int dE\;\Omega(E)e^{-\beta N_p E}.
\end{equation}
This is done using the Ferrenberg-Swendsen algorithm \cite{ferrenberg1989optimized}. Moving into the complex plane, the real and imaginary parts of the partition function will be
\begin{equation}
    \text{Re}(Z(\beta)) = \int dE \;\Omega (E) e^{-\text{Re}(\beta N_pE)} \cos(\text{Im}(\beta N_pE))
\end{equation}
and
\begin{equation}
    \text{Im}(Z(\beta)) = -\int dE \;\Omega (E) e^{-\text{Re}(\beta N_pE)} \sin(\text{Im}(\beta N_pE)).
\end{equation}
By scanning the complex $\beta$ plane, we construct the curves where the Real and Imaginary parts of $Z$ are zero. The complex zeros are then located at the intersections of these curves. 

\section{General scaling expectations}\label{scaling}
It seems empirically clear from Refs. \cite{Jin:2009mc, Jin:2011eon, Jin:2013hpa,gelzer2013fisherzerosrgflows, gelzer2014exploringphasestructure12flavor, e2018critical} that for low enough values of $m_q$, we have a first order phase transition manifested by a discontinuity in the average plaquette, and for $m_q$ large enough a smooth crossover. In these two limits, we have rather simple expectations from Renormalization group (RG) considerations. In the case of the Ising model at a fixed temperature and a variable external magnetic field $h$, the partition function at finite volume is proportional to finite polynomial in $\exp(-2h)$. In addition, the partition function is invariant under the change $h\rightarrow -h$. Lee and Yang \cite{leeyang} showed that the zeros in the $\exp(-2h)$ complex plane are located on the unit circle and pinch the real axis when $T<T_c$, while a gap remains for $T>T_c$. Later \cite{itzykson83}, these results were reformulated in terms of scaling variables and extended \cite{janke2000} to models where the field variables are continuous. 
These findings are summarized in \cite{janke2000} that we now follow for our problem. 

For $m_q<m_q^c$, there is a first-order phase transition and we expect that at infinite volume, the zeros pinch the real axis preventing an analytical continuation between the two phases. Using a generic model for the zeros of the partition function and the free energy density \cite{janke2000} for a linear size $L$ in $D$ dimension, it is possible to show that by assuming that the imaginary part of the lowest zeros decay like $L^{-D}$, a discontinuity is observed in the first derivative of the free energy density. On the other hand for a second-order phase transition, we expect the lowest zeros to decay like $L^{-1/\nu}$ and no discontinuity in the first derivative is observed if $\nu > 1/D$. However, a singularity in the second derivative will appear if $\nu \leq 2/D$, in agreement with the scaling of the specific heat exponent $\alpha=2-D\nu$.

In the case $m_q>m_c$, assuming a smooth crossover at the point where the second derivative of the average energy vanishes, we should be able to analytically continue across this point and a gap between the lowest zeros and the real axis should be present. Following Refs. \cite{fisher78,itzykson83}, we expect this gap to follow the scaling $(m_q-m_c)^{\beta \delta}$. The value of $\beta\delta$ is 3/2 for a mean field transition and 15/8 for the two dimensional Ising model.

\section{Results}
\label{allresults}
\subsection{Determination of the zeros}
\label{sec:results}
\begin{figure}[h]
    \centering
    \includegraphics[width=0.8\linewidth]{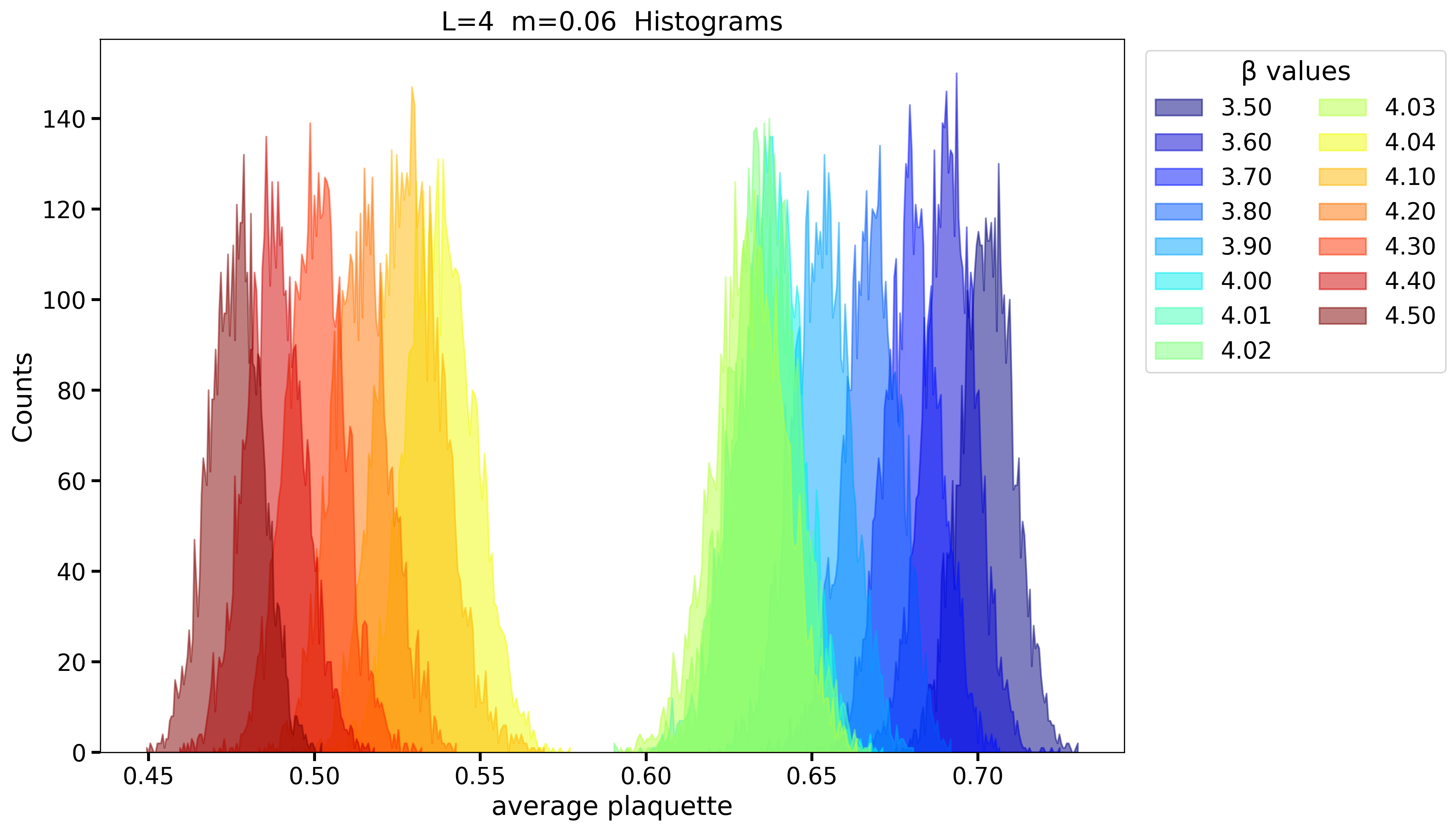}
    \caption{Distribution of average plaquette values for $L=4,\; m_q=0.06$ for $3.50\leq\beta\leq4.50$.}
    \label{fig:histograms}
\end{figure}

\begin{figure}[h]
    \centering
    \includegraphics[width=0.8\linewidth]{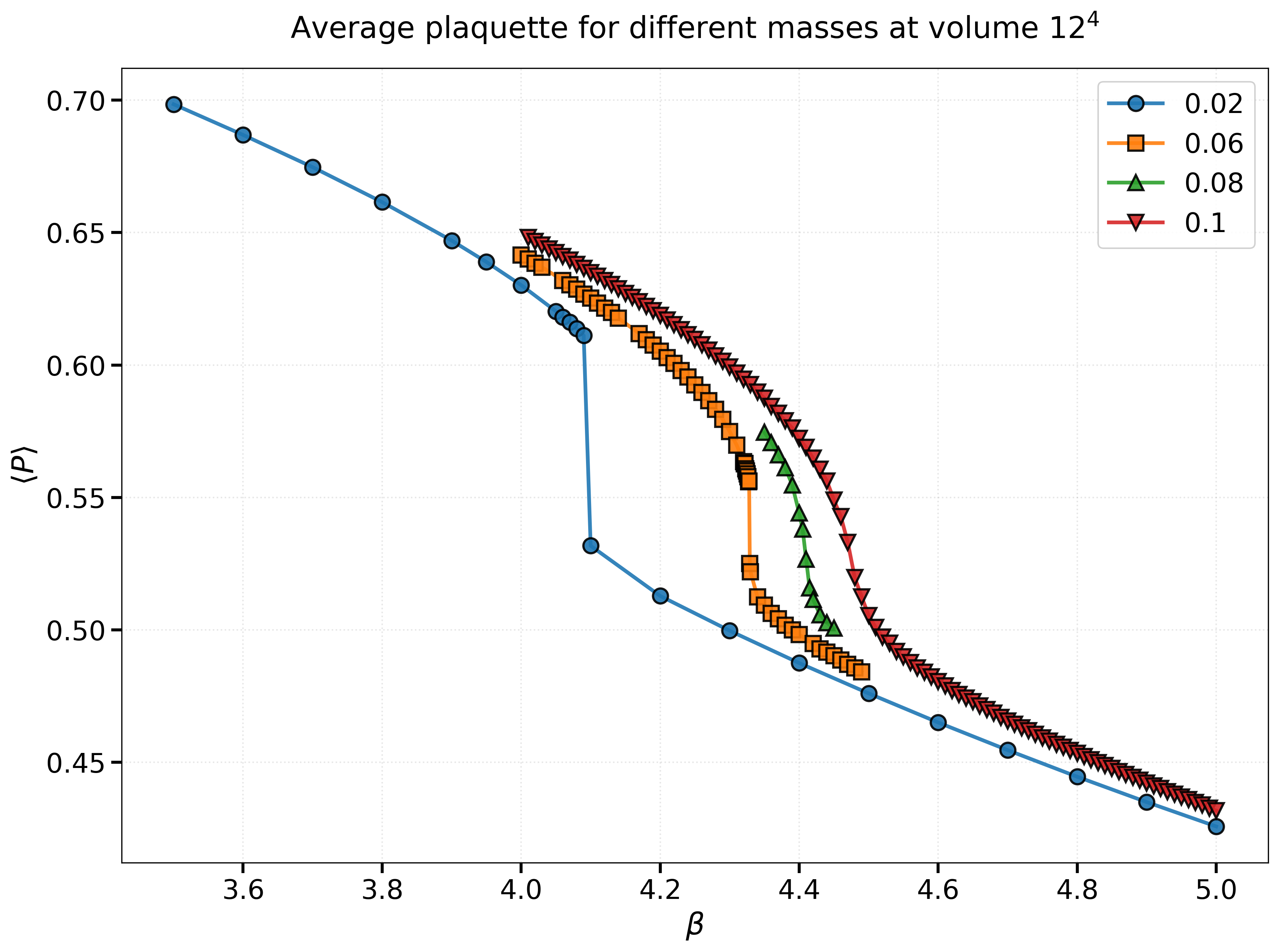}
    \caption{Distribution of average plaquette values as a function of $\beta$ for $L=12$ and masses $m_q=0.02,\;0.06,\;0.08,$ and $0.1$.}
    \label{fig:average-plaquette-value}
\end{figure}

The data was generated by the unimproved hybrid Monte Carlo (HMC) algorithm for masses $m_q=0.02, 0.06, 0.08$ and $0.1$ for a range of $\beta$ defined in terms of the bare gauge coupling $g$ as $\beta = 6/g^2$. More details are provided in Appendix \ref{app:data}. After thermalization, we use Wolff's bootstrapping method \cite{wolff2004monte} to determine the autocorrelation function and correlation times. As shown in Fig. \ref{fig:histograms}, the data is then binned for the various values of $\beta$ and passed through the Ferrenberg-Swendsen algorithm to numerically determine the density of states and the partition function. The procedure is discussed in Appendix \ref{app:algo}. The average plaquette for different values of $\beta$ are shown in Fig. \ref{fig:average-plaquette-value} for the four values of $m_q$. This suggests that the first order transition turns into a crossover as $m_q$ is increased. 

The zeros are located by evaluating the real and imaginary parts of the partition function, identifying the contours where $Re(Z)=0$ and $Im(Z)=0$, and computing their intersections.

We were able to measure the Fisher zeros for different masses and tabulated the results in \cref{tab:m002,tab:m006,tab:m008,tab:m01} in \cref{app:numerical results}. Uncertainty quantification was performed in two ways: (i) via bootstrapping analysis, and (ii) by estimating the error arising from shallow-angle intersections of the contours. Such shallow intersections can amplify small numerical fluctuations in the partition function leading to significant shifts in the inferred zero locations. This is illustrated in Fig. \ref{fig:contours}. Additional numerical details are provided in Appendix B and the supplementary material.

\begin{figure}
    \centering
    \includegraphics[width=1\linewidth]{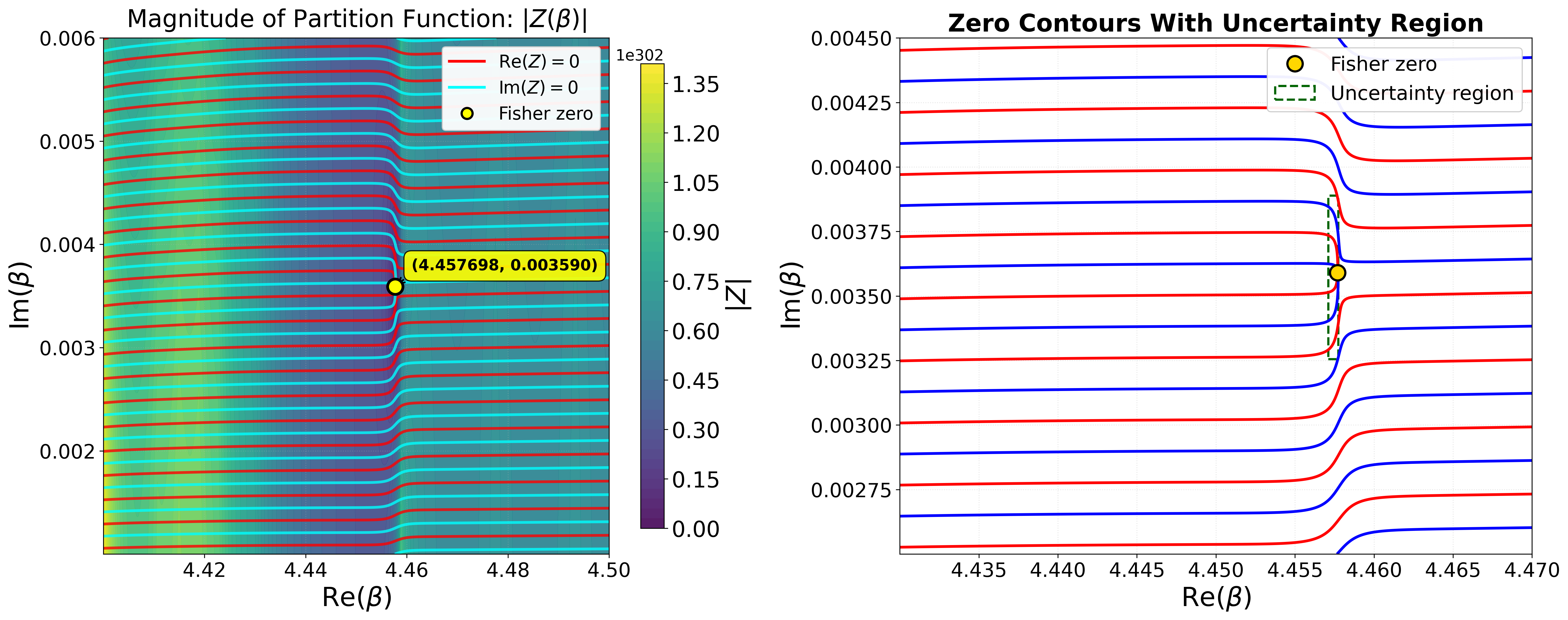}
    \caption{(Left) Heatmap of the of the magnitude of the partition function along with the real and imaginary zero contours for $L=8,\;m_q=0.08$. (Right) Zoomed in contour plot with an associated uncertainty region for the Fisher zero.}
    \label{fig:contours}
\end{figure}

\subsection{Fits of zeros as function of the size}
\begin{figure}
    \centering
    \includegraphics[width=1\linewidth]{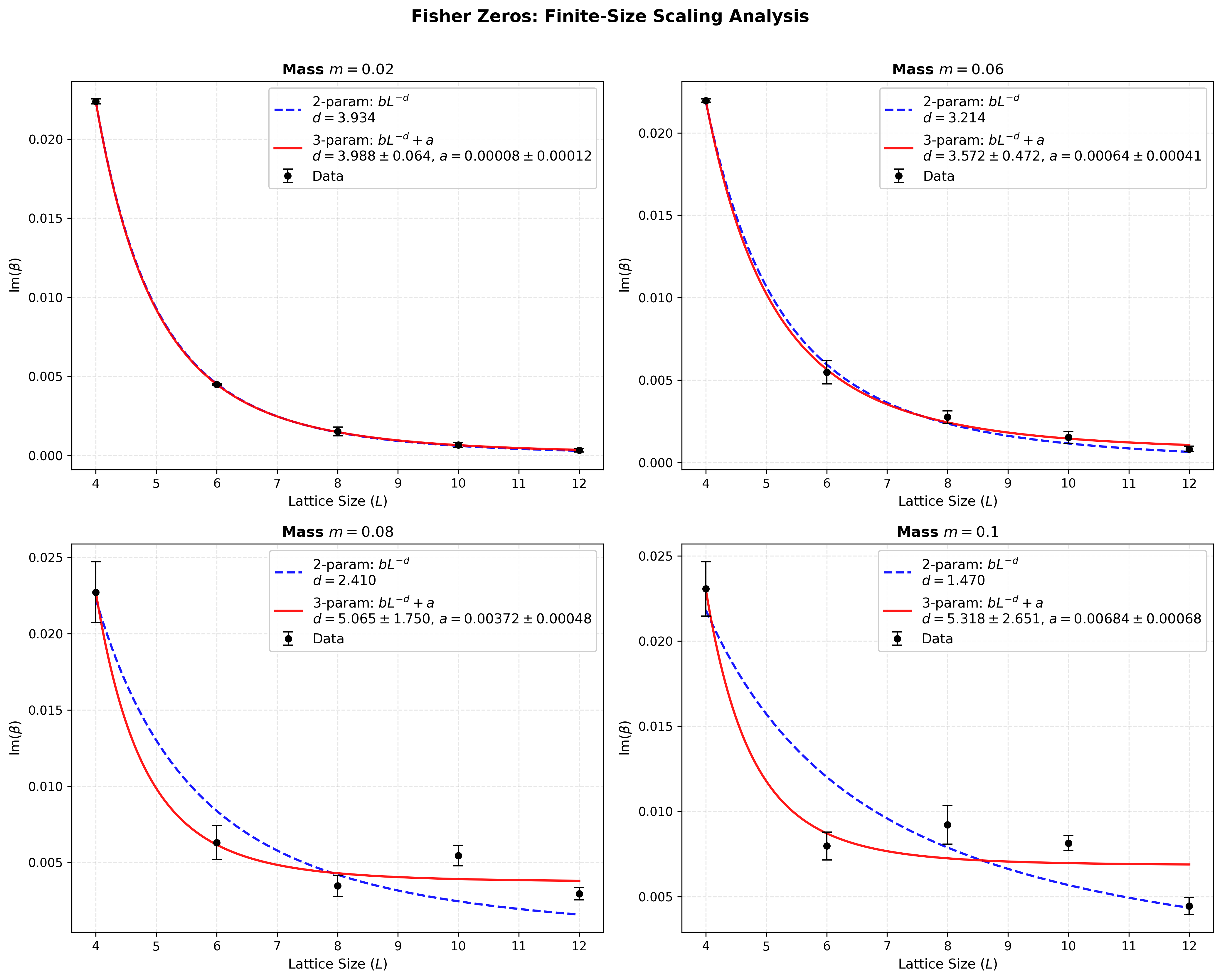}
    \caption{Imaginary part of the first Fisher zero as a function of the linear lattice size $L$ for $m=$0.02, 0.06, 0.08, 0.1.}
    \label{fig:scalings}
\end{figure}

We used two and three parameter fits of the form $y = bL^{-d}$ and $y = a + bL^{-d}$, respectively, where $y$ is the imaginary part of the Fisher zero and $L$ is the lattice size. The results are shown in Tables \ref{tab:par2} and \ref{tab:par3}, and Fig. \ref{fig:scalings}. The errors in the fitting parameters were calculated via bootstrap analysis using the errors evaluated for the zero locations. By doing this, we included both the statistical and systematic errors present in the zeros locations as discussed earlier. Furthermore, we evaluated the reduced $\chi^2$, defined as $\chi_{\text{reduced}}^2 = \frac{\chi^2}{\nu}$, where $\nu$ are the degrees of freedom, for both models. The results are provided in Table \ref{tab: chi}.

\begin{table}[h]
\centering
\caption{\label{tab:par2}The fitting parameter $d$ for different masses for the 2-parameter fits}
\begin{tabular}{c|c}
\hline
$m_q$ & $d$ \\ 
\hline
0.02  & 3.934 $\pm$ 0.066  \\
0.06  & 3.231$\pm$ 0.200  \\
0.08  & 2.411 $\pm$ 0.206  \\
0.1  & 1.467$\pm$0.142  \\
\hline
\end{tabular}
\end{table}

\begin{table}[h]
\centering
\caption{the fitting parameters $d$ and $a$ for different masses for the 3-parameter fits}
\begin{tabular}{c|c|c}
\hline
$m_q$ & $d$ & $a$\\ 
\hline
0.02  & $3.988\pm 0.064$  & $0.000076 \pm 0.000120$\\
0.06  & $3.572 \pm 0.472$ & $0.000637 \pm 0.000409$ \\
0.08  & $5.065 \pm 1.750$ & $0.003722 \pm 0.000484$\\
0.1  & $5.318 \pm 2.651$ & $0.006837 \pm 0.000682$  \\
\hline
\end{tabular}
\label{tab:par3}
\end{table}

\begin{table}[h]
\centering
\caption{the $\chi^2$ and reduced $\chi^2$ for both models}
\begin{tabular}{c|c|c|c|c}
\hline
$m_q$ & 2-parameter $\chi_{\text{reduced}}^2$ & 2-parameter $\chi^2$ & 3-Parameter $\chi_{\text{reduced}}^2$  & 3-Parameter $\chi^2$ \\ 
\hline
0.02  & $0.999$ & $2.998$  & $0.083$ & $0.166$\\
0.06  & $1.358$ & $4.075$ & $1.479$  & $2.958$ \\
0.08  & $11.762$ &  $35.286$ &$5.432$ &$10.865$\\
0.1  & $19.098$ &  $57.293$ &$17.343$ &$34.685$  \\
\hline
\end{tabular}
\label{tab: chi}
\end{table}

\begin{table}[h]
\centering
\caption{the $p$-value for both models}
\begin{tabular}{c|c|c}
\hline
Mass & 2-Parameter Fit & 3-Parameter Fit \\ 
\hline
0.02  & $0.3919$  & $0.9203$\\
0.06  & $0.2534$ & $0.2278$ \\
0.08  & $1.06\times10^{-7}$ & $4.37\times10^{-3}$\\
0.1  & $2.23\times10^{-12}$ & $2.94\times10^{-8}$  \\
\hline
\end{tabular}
\label{tab: p-value}
\end{table}

We now discuss the fits. The two parameter fit provides strong support for the hypothesis of a first order phase transition with a $L^{-D}$ (for $D$=4) approach of the real axis for $m_q=0.02$. As we increase $m_q$, the exponent decreases down to values that at first sight could be interpreted as close to 2 which is $1/\nu$ for a mean field transition. However, for $m_q$= 0.08 and 0.1, the $\chi^2$ values are very large and  correspond to a very small $p$-value. Fig. \ref{fig:scalings} suggests that for these two masses, the low quality of the fits may be due to the fact that the zeros do not pinch the real axis and that a gap subsists as $L$ goes to infinity. This possibility is allowed by the three-parameter fits which assume a possible non-zero asymptotic value for the imaginary part of the zeros denoted $a$. For $m_q=0.02$, we found $a=0.000076\pm 0.00012$, a value much smaller than the smallest zero for this mass and statistically compatible with 0. At the same time, the exponent gets closer to 4 with a better $p$-value which reinforces our first order transition hypothesis. For $m=0.06$, we found $a=0.00064 \pm 0.0004$ which is only slightly below the lowest zero. The fit has an acceptable $p$-value and we believe that $m_q=0.06$ is very slightly above the critical value $m_q^c$. This question is discussed in more detail in Sec. \ref{sec:mc}. For $m_q$= 0.08 and 0.1, significantly larger values of $a$ are found. Compared to the 2-parameter fits, the $\chi-$squares are significantly smaller and the $p$-values increase. However there is room for improvement. The low $p$-values can be explained from the fact that the imaginary parts of the zeros do not decay monotonically as seen in Fig. \ref{fig:scalings}. This indicates that higher order finite size scaling corrections are present and that we are significantly above the critical mass and that higher order finite size scaling is necessary. 

\subsection{Estimation of $m_q^c$ }
\label{sec:mc}
\begin{figure}[h]
    \centering
    \includegraphics[width=0.7\linewidth]{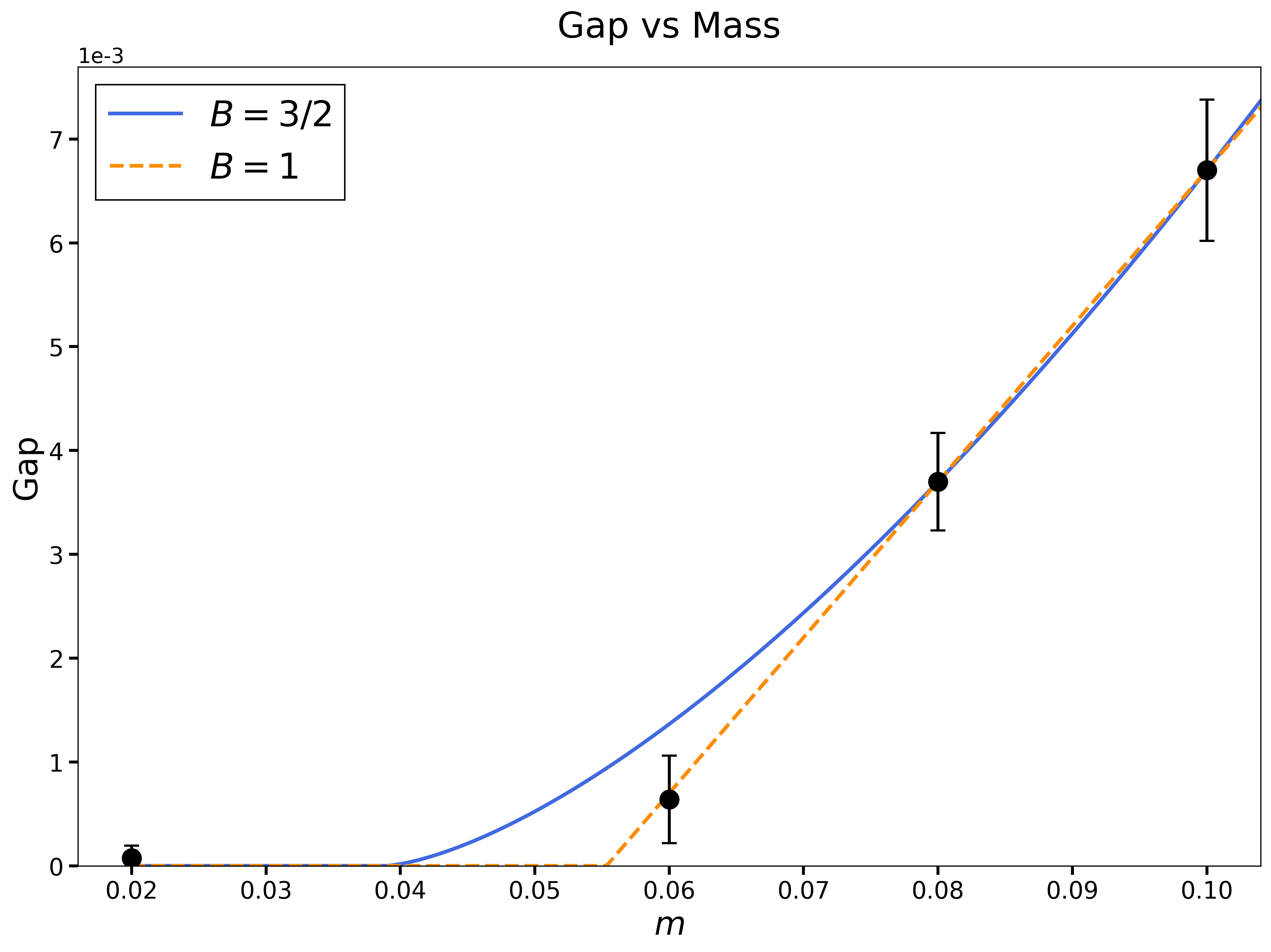}
    \caption    {Two models for the gap.}
    \label{fig:gap}
\end{figure}

We now plot in Fig. \ref{fig:gap} the asymptotic estimate of the zero gap at infinite volume, denoted  $a$ and provided in Tab. \ref{tab:par3},  versus $m_q$. We considered the following parametrization
\begin{equation*}
 a(m_q)=A(m_q-m_q^c)^B   
\end{equation*}
Given that we have only three data points, we fixed $B$ at value 3/2 (mean field) or 1 (linear model) and used the results at $m=0.08$ and $0.1$ to determine $m_c$ and $A$. This predicts the value of $a$ for $m_q=0.06$ which can be compared with the numerical data and provide an indication about the quality of the fit. Numerically, we found $m_q^c=0.039$ and $A=0.44$ for $B=3/2$,  and $m_q^c=0.055$ and $A=0.15$ for $B=1$. It is visually clear that the linear fit predicts the value at $m_q=0.06$ much better than for $B=3/2$. This aligns with Klinger et al. \cite{klinger2026phasestructuremasslessmanyflavour} when there results for $N_f\leq 10$ are extrapolated to $N_f=12$ giving a critical mass of $m_q^c=0.0625$. However, the discrepancy for the prediction $B=3/2$ is less than two standard deviations and a more detailed  sampling of $m_q$ close to 0.05 should be conducted to determine $B$ more accurately. 

\section{Conclusion}
In summary, we considered a four-dimensional $SU(3)$ lattice gauge theory with 12 staggered fermions having identical masses and an unimproved action. Using sets of plaquette distributions for various inverse bare couplings $\beta$, we reconstructed the density of states with the Ferrenberg-Swendsen method and calculated the zeros of the partition in the complex $\beta$ plane with bare quark masses $m_q$ = 0.02, 0.06, 0.08 and 0.1 for hypercubes of linear size $L$= 4, 6, 8, 10, and 12. We fitted the $L$ dependence of the zeros with the lowest imaginary part using two ($y = bL^{-d}$) and  three ($y = a + bL^{-d}$) parameter fits. Our goal was to test the hypothesis that there is a line of first order transitions in the $(m_q,\beta)$ plane ending at a second order phase transition possibly in the 4D Ising (mean field) universality class. 

For $m_q$ = 0.02 the results provide strong support for a first order phase transition ($d\simeq 4$ and $a\simeq 0$). The results indicate with less statistical significance that the three other masses are above the critical value $m_q^c$ and that $a\sim (m_q-m_q^c)^{1.0}$. Combining this result with spectroscopic results by Jin and Mawhinney \cite{Jin:2009mc,Jin:2011eon,Jin:2013hpa}, indicate that the gap with the real axis (Lee-Yang edge) scales approximately like $m_\sigma ^2$, where $m_\sigma $ is the mass of the $0^{++}$ scalar which is also the lowest excitation. A more accurate description of the critical point could be obtained by scanning the quark mass more finely near $m_q\simeq 0.05$ and developing a RG picture beyond the linear approximation. 

\section*{Acknowledgments}
The authors thank Donald Sinclair and Yuzhi Liu for their help in the early stages of the project. We also thank Paul Zelle for comments on the manuscript. This work is supported in part by the Department of Energy under Award DE-SC0010113. MH was also funded in part by NSF award DMR-1747426. 


\appendix
\section{Data Structure and Parsing}\label{app:data}
We briefly describe the data structure. We considered volumes $L^4$ for $L=4,6,8,10,12$ and masses $m=0.02,0.06,0.08$ and $0.1$, and Monte Carlo sampled each case for a range of $\beta$'s given in Table \ref{tab:beta-table}. The number of Monte Carlo samples $N$ for a given $(L,m,\beta)$ ranged between 5,000 and 9,000 samples. The individual data (text) files \verb|data_0i| (\verb|i| $\in 1,\dots,5$) contain measurements of $N/5$ samples, with each sample having 14 different measurements. We are only interested in the average plaquette value \verb|APQ| for a given configuration, so we parse that quantity from each of the $(L,m,\beta)$ datasets. Doing this for each value of $\beta$, we end up with a length $N_\beta$ array containing lists of average plaquette values for a given $\beta$. This array is what is used in the analysis.
\begin{table}[h]
    \centering
    \caption{$\beta$ ranges for data of a given length $L$ and mass $m_q$. $\beta$ values in a range are not equally spaced, with more $\beta$'s taken around expected phase transitions.}
    \begin{tabular}{c|c|c|c|c}
        \hline
         $L$ & \;$m=0.02$\; & 0.06 & 0.08 & 0.1  \\\hline
         4 & [3.5, 4.5] & [3.5, 4.5] & [3.5, 4.5] & [4.0, 5.0]\\
         6 & [3.0, 5.0] & [4.05, 4.3]& [4.0, 5.0] & [4.0, 5.0]\\
         8 & [3.5, 4.5] & [4.0, 5.0] & [4.0, 5.0] & [4.0, 5.0]\\
         10 & [3.8, 4.2] & [3.8, 4.99] & [4.3, 4.41] & [4.2, 4.6]\\
         12 & [3.5, 5.0] & [4.0, 4.49] & [4.35, 4.45] & [4.01, 5.00]\\\hline
    \end{tabular}
    \label{tab:beta-table}
\end{table}

\section{Details of the numerical results}\label{app:numerical results}
The Fisher zeros we found are listed in the tables \cref{tab:m002,tab:m006,tab:m008,tab:m01}.

\begin{table}[h]
\centering
\caption{Fisher zeros for different lattice sizes at $m=0.02$}
\begin{tabular}{c|c}
\hline
Lattice Size & Imaginary part of the Fisher zero \\
\hline
4  & $0.022371\pm 0.000160$     \\
6  & $0.004488\pm 0.000031$    \\
8  & $0.001513 \pm 0.000280$    \\
10 & $0.000657 \pm 0.000162$    \\
12 & $0.000335 \pm 0.000106$    \\
\hline
\end{tabular}
\label{tab:m002}
\end{table}

\begin{table}[h]
\centering
\caption{Fisher zeros for different lattice sizes at $m=0.06$}
\begin{tabular}{c|c}
\hline
Lattice Size & Imaginary part of the Fisher zero \\
\hline
4  & $0.021958\pm 0.000100$     \\
6  & $0.005473\pm 0.000700$    \\
8  & $0.002748\pm 0.000370$    \\
10 & $0.001521\pm 0.000365$    \\
12 & $0.000814\pm 0.000168$    \\
\hline
\end{tabular}
\label{tab:m006}
\end{table}

\begin{table}[h]
\centering
\caption{Fisher zeros for different lattice sizes at $m=0.08$}
\begin{tabular}{c|c}
\hline
Lattice Size & Imaginary part of the Fisher zero \\
\hline
4  & $0.022725 \pm 0.002000$    \\
6  & $0.006295 \pm 0.001110$    \\
8  & $0.003466 \pm 0.000696$    \\
10 & $0.005449 \pm 0.000680$    \\
12 & $0.002945 \pm 0.000410$    \\
\hline
\end{tabular}
\label{tab:m008}
\end{table}

\begin{table}[h]
\centering
\caption{Fisher zeros for different lattice sizes at $m=0.1$}
\begin{tabular}{c|c}
\hline
Lattice Size & Imaginary part of the Fisher zero \\
\hline
4  & $0.023057 \pm 0.001600$    \\
6  & $0.007971 \pm 0.000820$    \\
8  & $0.009215 \pm 0.001140$    \\
10 & $0.008135 \pm 0.000443$    \\
12 & $0.004442 \pm 0.000500$    \\
\hline
\end{tabular}
\label{tab:m01}
\end{table}

We also looked at $\beta_C$ by finding the inflection points in the average plaquette vs $\beta$ graphs and you can see the results in Fig. \ref{fig:average-plaquette-value}

\begin{figure}
    \centering
    \includegraphics[width=0.5\linewidth]{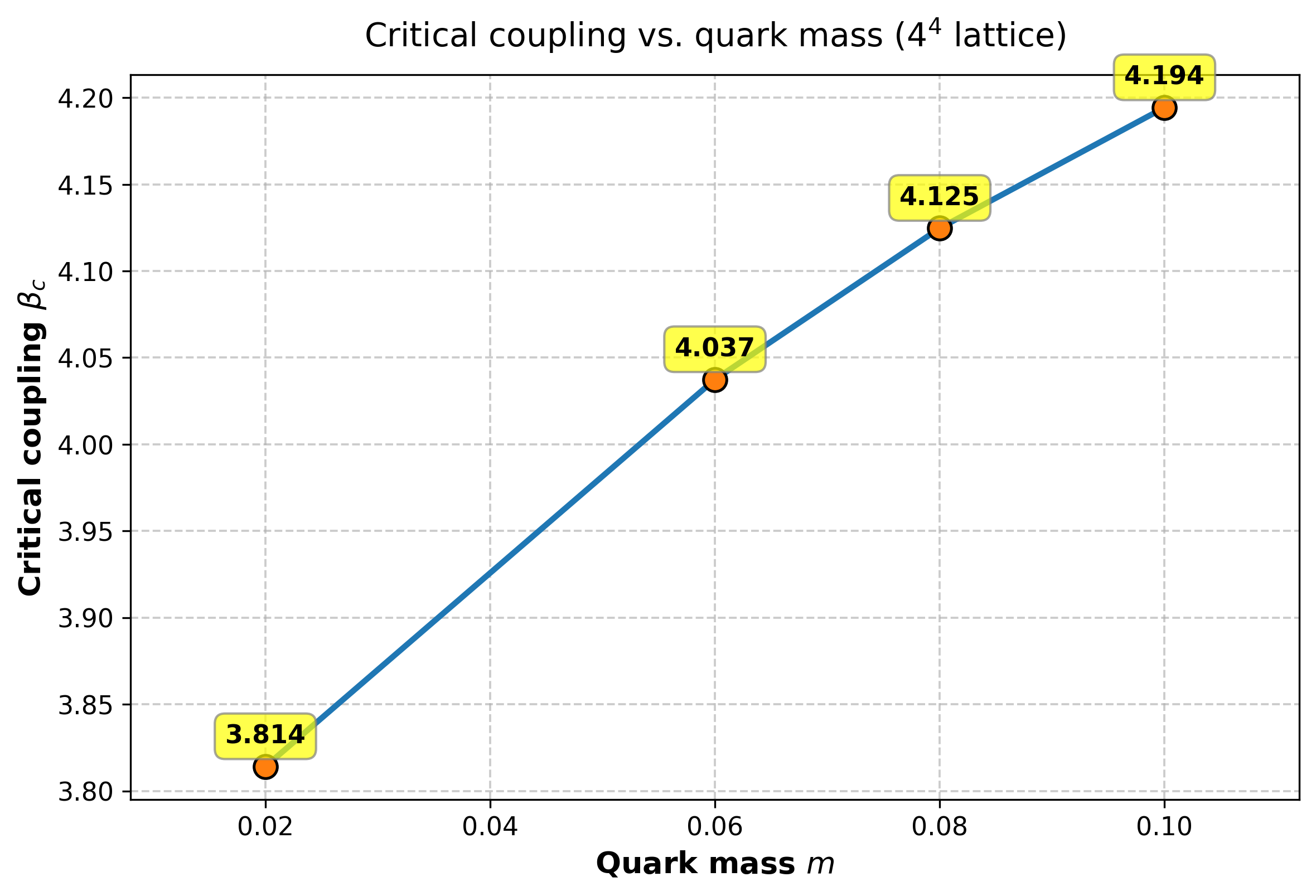}\includegraphics[width=0.5\linewidth]{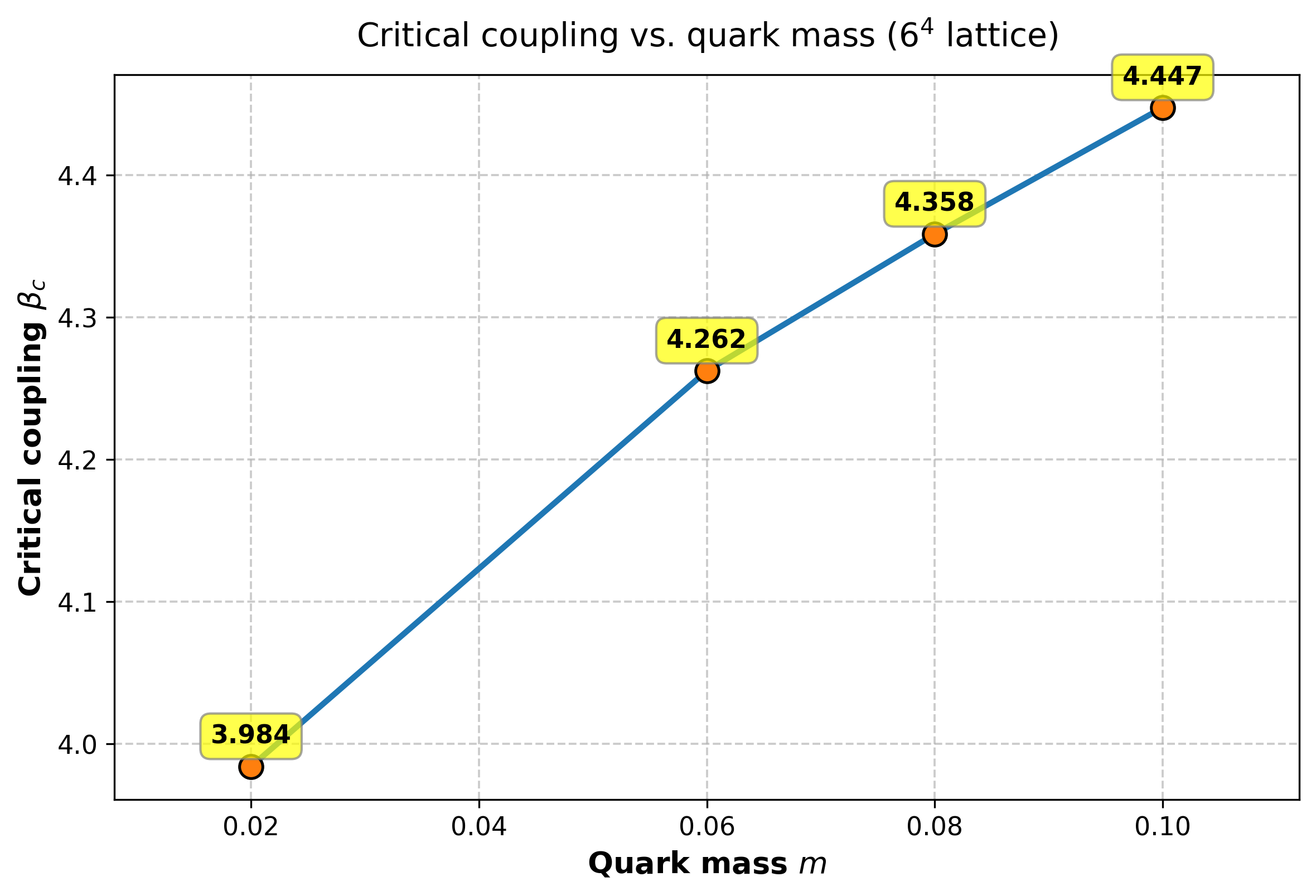}\\
    \includegraphics[width=0.5\linewidth]{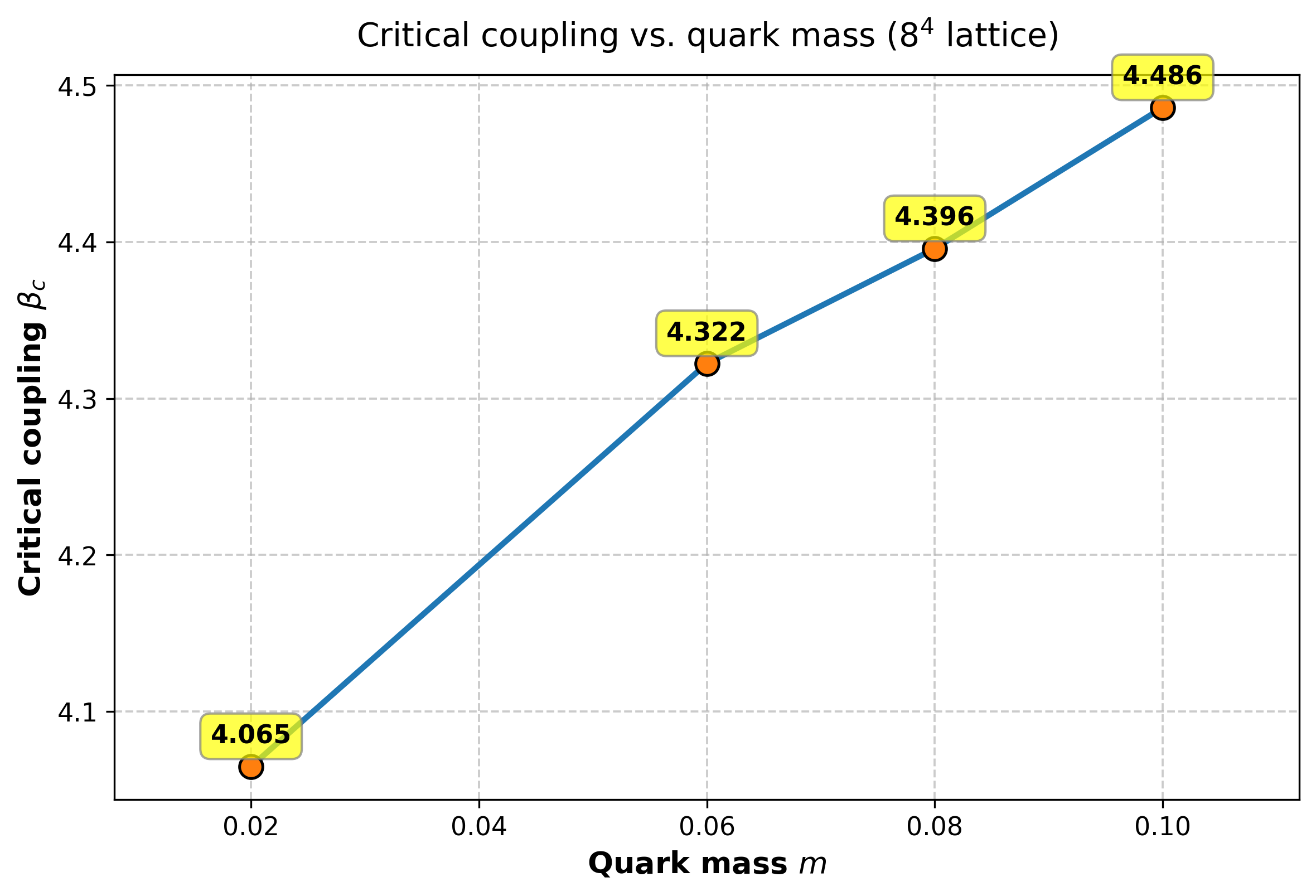}\includegraphics[width=0.5\linewidth]{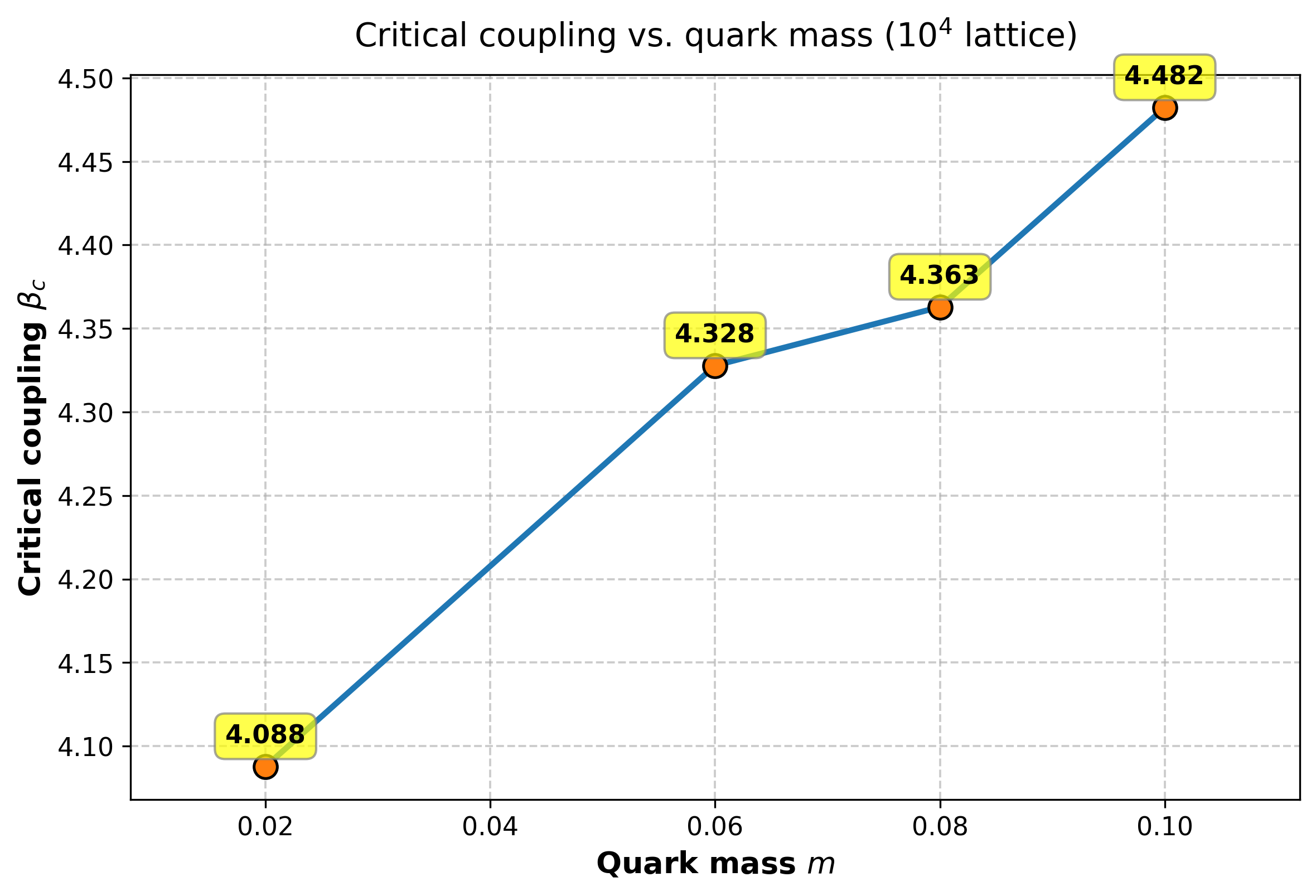}\\
    \includegraphics[width=0.5\linewidth]{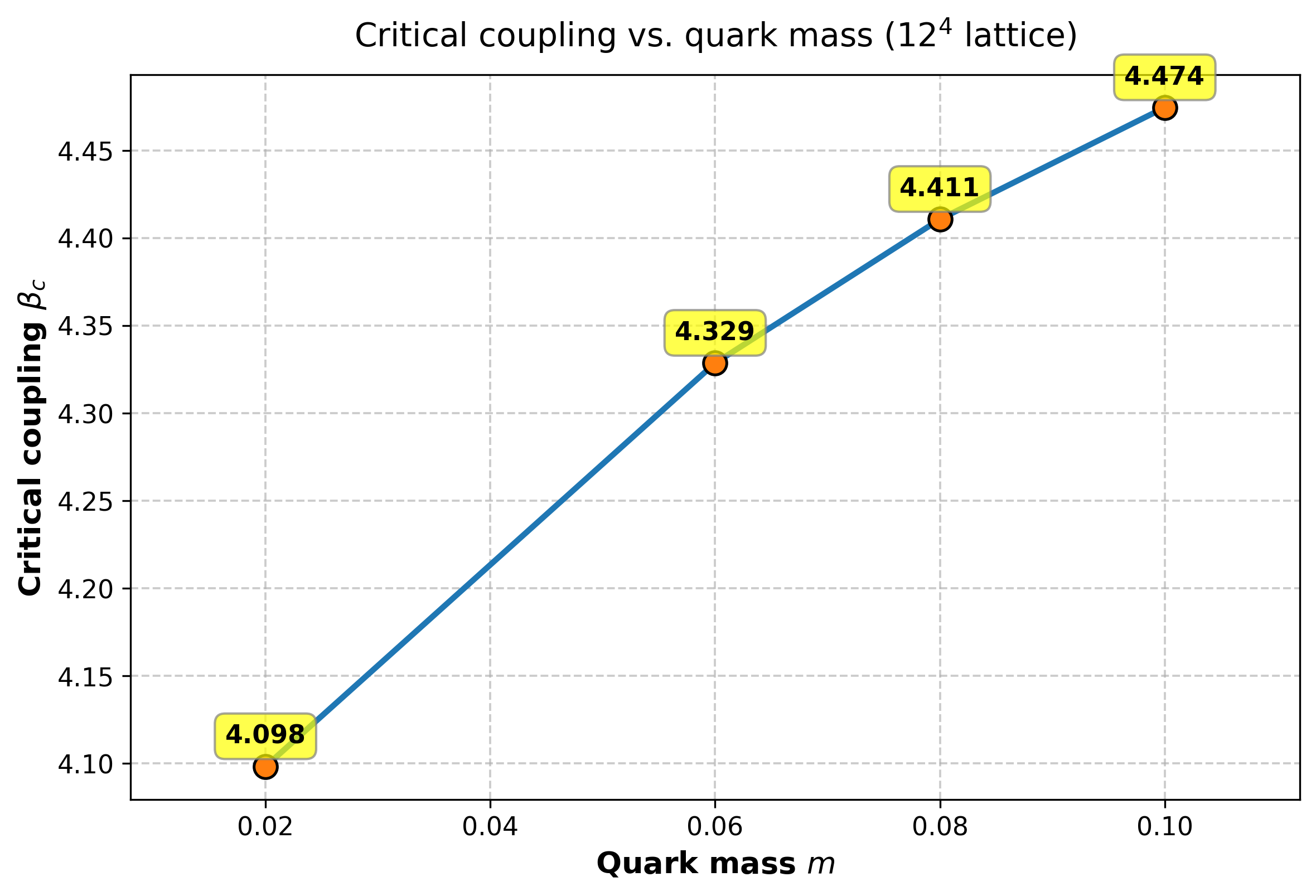}
    \caption{Critical coupling $\beta_c$ as a function of mass $m_c$ for $L=4, 6, 8, 10, 12$.}
    \label{fig:placeholder}
\end{figure}

\section{Computational Implementation of Bootstrap Fisher Zero Estimation}
\label{app:algo}
The analysis pipeline is implemented in Python. It begins by loading plaquette time-series data from the Monte Carlo simulation. After loading, the thermalization period is dropped from each simulation data(the exact number of points dropped varies by simulation and can be configured in the code). After that, we compute the autocorrelation function using the formula
\begin{equation}
\phi(t) = \frac{E[(X_i - \bar{X}) (X_{i+t} -\bar{X})]}{E[(X_i-\bar{X})^2]},
\end{equation}
and the correlation time using Wolff's method \cite{wolff2004monte}.

For each bootstrap sample, a block bootstrap is performed using block sizes 2$\tau$ to preserve temporal correlations in the Monte Carlo chain. Each bootstrap sample is packaged with its associated $\beta$ values, resampled plaquette data, correlation times, and lattice metadata. These samples are then processed independently.

Each bootstrap sample goes through the Ferrenberg–Swendsen algorithm\cite{ferrenberg1989optimized}, where we iteratively find the free energy of the system using
\begin{equation}
f_i^{\text{new}} = -\ln\left[ \frac{\sum_E H(E) \exp\left(-\beta_i N_p E - f_i^{\text{old}}\right)}{\left(\sum_j n_j^{\text{eff}} \exp\left(-\beta_j N_p E - f_j^{\text{old}}\right)\right)}\right],
\end{equation}
where
\begin{align*}
f_i &= \text{free energy at simulation } i, \\
\beta_i &= \text{inverse coupling parameter for simulation } i, \\
N_p &= 6L^4 \text{ (number of plaquettes)}, \\
E &= \text{plaquette observable value}, \\
H(E) &= \text{combined histogram at plaquette } E, \\
&\quad \text{weighted by } g_j = 1+2\tau_j, \\
n_j^{\text{eff}} &= \text{effective number of measurements for simulation } j.
\end{align*}
For numerical stability we do this in log-space
\begin{align*}
f_i^{\text{new}} &= -\text{LSE}_E\bigg[\ln\left(H(E)\right) - \beta_i N_p E \notag \\
&\quad + \text{LSE}_j\left[\ln n_j^{\text{eff}} - \beta_j N_p E - f_j^{\text{old}}\right]\bigg]    
\end{align*}
where LSE(LogSumExp) is defined as
\begin{equation}
\text{LSE}[x_k] =x_{\max} + \ln\left(\sum_k e^{x_k - x_{\max}}\right)
\end{equation}
where $x_{\max} = \max_k(x_k)$.
Once the free energy difference is less than $10^{-8}$ or after the 10,000th iteration, we use the final free energy and calculate the density of states:
\begin{equation}
\Omega(E) = \frac{H(E)}{\sum_j n_j^{\text{eff}} \exp\left[-\beta_j N_p E - f_j\right]},
\end{equation}
in log-space as
\begin{equation}
\ln \Omega(E) = \ln H(E) - \text{LSE}_j\left[\ln n_j^{\text{eff}} - \beta_j N_p E - f_j\right]
\end{equation}

Then we evaluate the complex partition function, defined as 
\begin{equation}
Z(\beta) = \int \Omega(E) \exp[-\beta N_p E] \, dE
\end{equation}
on a grid in the complex $\beta$ plane. We approximated this integral as a sum over the density of states bins, which were previously evaluated.

To find the Fisher zeros, we started by finding the zero contours of both the real and imaginary part of the partition function using "skimage.measure.findcontours" \cite{lorensen1998marching}. After that, we performed a KD-tree search \cite{bentley1975multidimensional} between the list of points in the $Re(Z)=0$ contour and the $Im(Z)=0$ contour where we save any two points with a distance between them of less than some adaptive tolerance. We use these points as initial starting points for "scipy.optimize.fsolve" to pinpoint the zeros of the partition function, and finally, we remove duplicates by checking if any two zeros are within a distance of $10^{-6}$ of each other.

In cases where multiple zeros are identified, only the zero with the smallest imaginary part is retained; the others are discarded. The results reported in the text represent the mean values of these selected zeros, along with associated uncertainties from two sources. (The first is a statistical error, estimated via bootstrapping: we constructed a 68\% confidence interval around the mean to quantify this contribution. The second is a systematic error caused by the intersection between the real and imaginary contours at extremely shallow angles, which makes the exact zero location highly sensitive to numerical accuracy and any small changes in the partition function values. This error was estimated by constructing a set of all the points around the intersection with a closest distance between the contours smaller than the regular spacing of the contours.

\clearpage
\bibliography{apssamp}

\end{document}